\documentclass[prd]{revtex4}
\usepackage{graphicx}
\usepackage{dcolumn}
\usepackage{bm}
\usepackage{epsfig}
\begin{document}

\title{Arbitrary black-string deformations in the black string-black hole transitions}

\author{Matthew Anderson$^{1,2}$, Luis Lehner$^{1}$, Jorge Pullin$^{1,2}$}
\

\affiliation{$1$ Department of Physics and Astronomy, Louisiana State
University, Baton Rouge, LA 70803-4001\\
$2$ Center for Computation and Technology, Louisiana State
University, Baton Rouge, LA 70803-4001 }

\begin{abstract}
  We study the possible black string-black hole transition by
  analyzing the structure of the apparent horizon for a large family
  of time-symmetric initial data. We observe that, as judged by the
  apparent horizon, it is possible to generate arbitrarily deformed
  black strings at a moment of time symmetry. A similar study for
  hyperspherical black holes reveals that although arbitrarily
  deformed hyperspherical black holes can be constructed, the proper
  distance between the north and south poles along the extra direction
  has an upper limit.
\end{abstract}
%\pacs{}
\maketitle

\section{Introduction}
The dynamics of black strings has been the subject of significant
attention in recent years and several results have appeared in the
literature indicating the evolution of a sufficiently long perturbed
blackstring is likely richer than anticipated.  Considerable
interest was initially 
sparked when Gregory and Laflamme \cite{GL,GL2}, noted that in linearized
perturbation theory black strings longer than a certain length become
unstable. The work of Horowitz and Maeda shows that if the spacetime
continues to be asymptotically predictable, black strings cannot pinch-off 
in finite affine time \cite{Gary} and also presents additional
arguments against the pinch-off possibility in lieu of the perturbed
string evolving towards a stationary non-uniform string.  This
suggestion has spurred several papers ---and conjectures--- about what
fate a perturbed string might encounter (see for instance
\cite{Gary,kolreview,harmarkreview}).

However, there is not yet agreement on what the final fate might be
and different analyses are being carried out to shed light on this
issue.  For instance, stationary solutions which are non-uniform in
the extra dimension have been obtained \cite{Toby,gubser} but their
entropy (for dimensions lower than $14$ \cite{kolsorkin}) is too large
for these solutions to be the end point of the dynamical evolution
envisaged by Horowitz and Maeda\cite{Toby,gubser}.  Further analysis
of highly distorted stationary strings have revealed the spacetime
around the neck approaches one having a conical structure. It is
therefore conjectured that a perturbed string can pinch-off
discontinuously, experiencing a conical phase\cite{kol1,kol2}.

Numerical simulations have also been presented which reveal the rich
dynamics of a perturbed black string, where the initially
exponentially fast varying geometry characterizing the (highly
distorted) horizon neighborhood gives rise to a later stage where the
dynamics are significantly slowed down \cite{Simulation}. Although lack
of resolution prevented pushing the simulation further, a study of the
late stages of the solution indicated that the possibility of the
horizon pinching off in infinite affine time is quite
possible \cite{garfinkle}.  Extrapolating the observed results further,
the pinching off, if it happens, could proceed in a smooth way. Though
this possibility, as well as all other conjectures, awaits for numerical 
evolutions for a definitive statement.

There is still much to be understood before a clear picture of the
problem emerges.  It turns out interesting information can be
extracted by considering a sequence of initial data which aims to
probe a region close to a bifurcation scenario. This line of work has
been previously pursued by Sorkin and Piran \cite{sorkinpiran}
inspired by related work in the study of binary black hole collisions.
At the core of the idea is to find particular solutions at a given
instant of time (thus satisfying the constraint equations) which
describe a highly distorted black string (as judged by their apparent
horizons) and examine what occurs as this distortion is made more
extreme. The constraints are solved in the presence of a matter source
with at least one free parameter characterizing its size. By varying
this parameter, one can find hyperspherical black holes ---when the
source is sufficiently compact and localized--- or black strings
---when the source is compact but distributed along the extra
dimension.  By fine-tuning this parameter one can thus study what
type of transition is observed in highly distorted cases. To simplify
the task of finding consistent data in the presence of this source a
moment of time symmetry can be chosen, and so the momentum constraint
is trivially satisfied. The solution of the Hamiltonian constraint
yields the data sought after.  Once this is found, an analysis of the
apparent horizon at that hypersurface gives an indication of what the
behavior may be.  Unfortunately, this approach cannot give the event
horizon behavior, but since it must contain the apparent horizon, the
latter does give some indication of how the former behaves.

In the present work we study the problem and obtain numerical
solutions for a variety of sources which are chosen to probe deeply
the transition region between black holes and strings.  We consider a
$D$ dimensional spacetime endowed with a $SO(D-2)$ symmetry and
develop a code able to yield solutions of the initial data problem for
Einstein's equations at a moment of time symmetry in the presence of
an arbitrary chosen source. This allows us to probe a family of
solutions interpolating between hyperspherical black holes and black
strings.  To do so, we examine a sequence of solutions describing
distorted black holes and strings and study their properties. We
define such a sequence by varying the source while keeping the two
relevant scales in the problem ---namely the ADM mass per unit length 
of the spacetime and the asymptotic length of the extra 
dimension--- constant. This
fact, coupled to the ability of examining arbitrary spherical
dimensions and the use of computational techniques that let us probe
significantly distorted strings are the main differences with the work
presented in \cite{sorkinpiran}. As we will see, our
results agree with most of those presented there, but also interesting
differences arise.  Most importantly, we are able to produce
arbitrarily distorted strings by appropriately varying the source
considered. By analyzing these strings, we find features common to the
solutions found in simulations of perturbed black strings which could
explain the slow down of the dynamics observed there
\cite{Simulation}.

\section{Problem Set Up}

The analysis of allowed solutions in arbitrary dimensions has
recently revealed a possible interesting behavior at large enough
dimensions \cite{sorkin,kudohhigherd} through a linearized
analysis of the solution's behavior. Indeed these results would
indicate in dimensions $> 13$ the conjecture of Horowitz--Maeda, namely
that a perturbed string might settle down to a non-uniform stationary
string, is quite more plausible than in lower dimensional cases. This
is hinted by the behavior of thermodynamical quantities which
revealed the transition from a uniform string to a distorted one is
continuous for high enough dimensions.

It is thus interesting to consider finding solutions for arbitrary
dimensions and examine the possible differences that may arise. If
this marked transition at sufficiently high dimensions is robust
beyond the linear regime, its effects would likely be evident in the
solutions obtained. However, obtaining such solutions in large
dimensions is a delicate enterprise since fields near the origin
acquire a much steeper gradient. Furthermore, if one is interested in
constructing highly distorted situations ---in particular to discern
the behavior at a threshold of a very thin string--- numerical
techniques capable of probing these scenarios must be employed. We
thus adopt the following strategy to construct a code capable of
dealing with these issues.

We begin by extending the same approach followed in \cite{sorkinpiran}.  We
consider time-symmetric, space-like hypersurfaces and choose a
conformally flat metric for simplicity:
\begin{eqnarray}
dl^2 = \psi^2 \left( dr^2 + dz^2 + r^2 d\Omega^2_{D-3} \right),
\end{eqnarray}
where $D$ is the dimension of a spacetime with $SO(D-2)$ symmetry.
While the momentum constraint is trivially satisfied, the Hamiltonian
constraint becomes
\begin{eqnarray}
\partial_{rr} \psi + \partial_{zz} \psi 
+\frac{\left(D-3\right)}{r} \partial_r \psi
+ \left[ \left( \partial_r \psi \right)^2 
+ \left( \partial_z \psi \right)^2 \right]
\left[ \frac{\left( D-4 \right) \left( D-3 \right)}{2} 
- 1 \right]/\left(D-2\right)
                  + \rho \psi^3 = 0. \label{eqn:constraint}
\end{eqnarray}

Following the arguments given in \cite{yorksources}, a well posed
problem to solve for $\psi$ from Eq.~\ref{eqn:constraint} can be
obtained if the matter density is rescaled as $\rho \rightarrow
\tilde{\rho} \psi^{-3-s}$.

We solve the resulting equation with the use of finite element
techniques.  This enables a smooth mesh grading to support higher
resolution near the matter density.  Converting
Eq.~\ref{eqn:constraint} into the weak form for finite element
simulation also conveniently removes the coordinate singularity at
$r=0$.  Simulations used bilinear quadrilateral elements with 
full approximation scheme (FAS)
multigrid as the solver \cite{diffpack2}.  We used
the Diffpack finite element toolkit \cite{diffpack} in all development
and production runs.

In the present paper we concentrate on the $D=5$ case both for testing
purposes and to re-examine the results in \cite{sorkinpiran} with a
well defined sequence that keeps both the ADM mass and the asymptotic
length of the extra dimension fixed. We do so by appropriately varying
the source. The results for higher dimensions will be presented
elsewhere.

In five dimensions, Eq.~\ref{eqn:constraint} becomes
\begin{eqnarray}
\frac{1}{r^2} \partial_r \left( r^2 \partial_r \right ) \psi 
+ \partial_{zz} \psi + \tilde{\rho} \psi^{-s} = 0. 
\label{eqn:constraint5d}
\end{eqnarray}
We chose $s = 1$ for all simulations in the present case.  The non-physical 
density $\tilde{\rho}$ was chosen to be
\begin{eqnarray}
\tilde{\rho} = A \exp\left( -r^2/\sigma_r^2 \right) 
                  \left[ \exp\left( -z^2/\sigma_z^2 \right) 
+ \exp\left( -\left(z-2L\right)^2/\sigma_z^2 \right) \right]
              + A\exp\left(-r^2/\sigma_A^2\right) 
\exp\left(-z^2/\sigma_B^2\right), \label{eqn:source}
\end{eqnarray}
with $\sigma_r = 0.1$, $\sigma_z = 0.1$, and $\sigma_A \in
\left[0.025,1.0\right]$ and $\sigma_B \in \left[0.1,2.0\right]$.  The
first term of this source is also identical to reference
\cite{sorkinpiran}.  The second term of the source is intended to increase
the distortion of the black object horizon by increasing the matter
density in the periodic direction close to $r=0$.

In our sequence of solutions, the amplitude, $A$ was varied to ensure
the ADM mass remained the same for all density distributions.  The
size of the periodic dimension, $2L$, was chosen identical to
reference \cite{sorkinpiran} ($L=2$) and longer ($L=15$) so as to
cover cases above and below Gregory--Laflamme's critical length.
The adopted boundary conditions are as those in \cite{sorkinpiran}: reflection
symmetries, $\partial_r \psi = 0$ along $r=0$ and $\partial_z \psi =
0$ along $z=0$; periodicity, $\partial_z \psi = 0$ along $z=L$; and
Robin, $\partial_r \left( r \psi \right) = 1$, at the $r$ outer
boundary. The last condition ensures the asymptotic length of the
extra dimension is constant.
All our simulations have been carried out for different locations of the
outer boundary given by $r=5,10,20,50$ without observing significant
differences in the obtained solution.

Variations in the non-physical density distribution can produce either
black strings or hyperspherical black holes.  Locating the apparent
horizon indicates the type of black object present in the initial
data.  The apparent horizon is the outermost surface of zero-expansion
of null rays.  For initial data with vanishing extrinsic curvature,
the apparent horizon satisfies
\begin{eqnarray}
\nabla_i s^i = 0, \label{eqn:horizon}
\end{eqnarray}
where $s^i$ are the spatial indices of the space-like normal to the 
horizon surface.  For black string solutions, we consider the function
$\phi = r - \rho(z)$ and find the surface where $\phi = 0$ by 
solving Eq.~\ref{eqn:horizon}.  For black hole
solutions, we transform to spherical coordinates $R,\chi$ where 
$r=R \sin\left(\chi\right)$ and $z = R \cos\left(\chi\right)$, following
the approach described in \cite{sorkin}.  We then consider the function
$\phi = R - \rho(\chi)$ and locate the surface where $\phi = 0$.

\section{Results}

%The obtained code was first validated by considering several representative
%cases and monitoring the behavior of 

\begin{figure}
\begin{center}
\epsfig{file=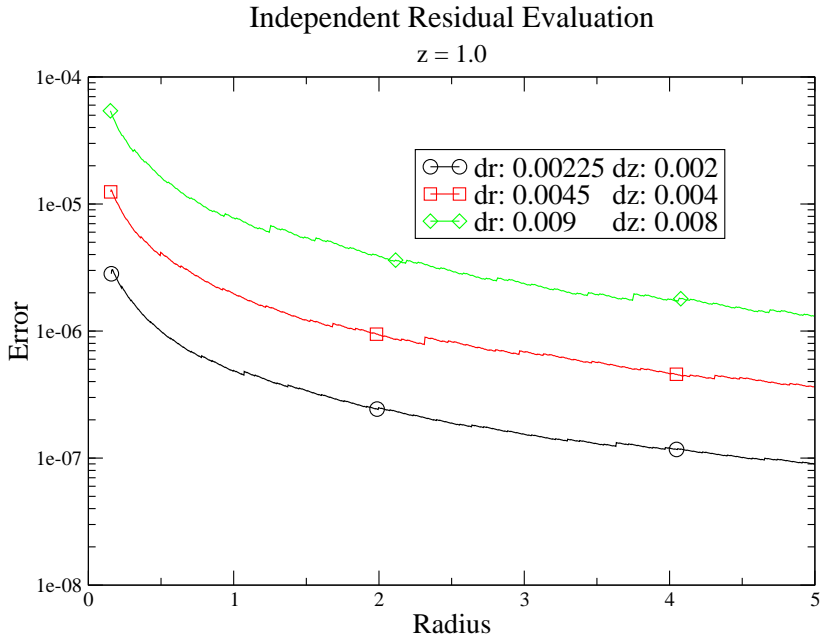,height=12.5cm,angle=270}
\caption{The unscaled error for a uniform black string along the 
radius at $z=1$  at three resolutions.  For the highest resolution 
at $r = 0.01$, the  error scaled by $\tilde{\rho}+1$ was 
$1.3 \times 10^{-5}$;  $\tilde{\rho}$ was $3.48 \times 10^5$.}
\label{fig:convergence}
\end{center}
\end{figure}

The code was tested using an independent residual evaluator.
Additionally, it was tested by recovering uniform black string and
hyperspherical black hole solutions.  Further tests were conducted
comparing cases using two entirely different finite element toolkits,
FlexPDE \cite{FlexPDE} and Diffpack \cite{diffpack}.  The tests cases
employed adaptive triangular meshes in the FlexPDE toolkit and graded
quadrilateral meshes in the Diffpack toolkit, with excellent agreement.
In all cases, good convergence of the solution was observed.  For
instance, Figure \ref{fig:convergence} shows the error as reported by
the independent residual evaluator for a uniform black string along
the radius at $z=1$ at three resolutions.  The convergence in the L2
norm is slightly better than first order.  As expected, the absolute
error is larger near the artificial matter distribution. A sense of the relative
error with respect to the defined source 
is given by monitoring the error divided by $\tilde{\rho}+1$.
All results apart from the convergence tests used resolutions 
of $dr = 0.00225$ and $dz = 0.002$, unless otherwise noted.

After validating the code, we proceeded to study both distorted strings
and hyperspherical black holes.

\begin{figure}
\begin{center}
\epsfig{file=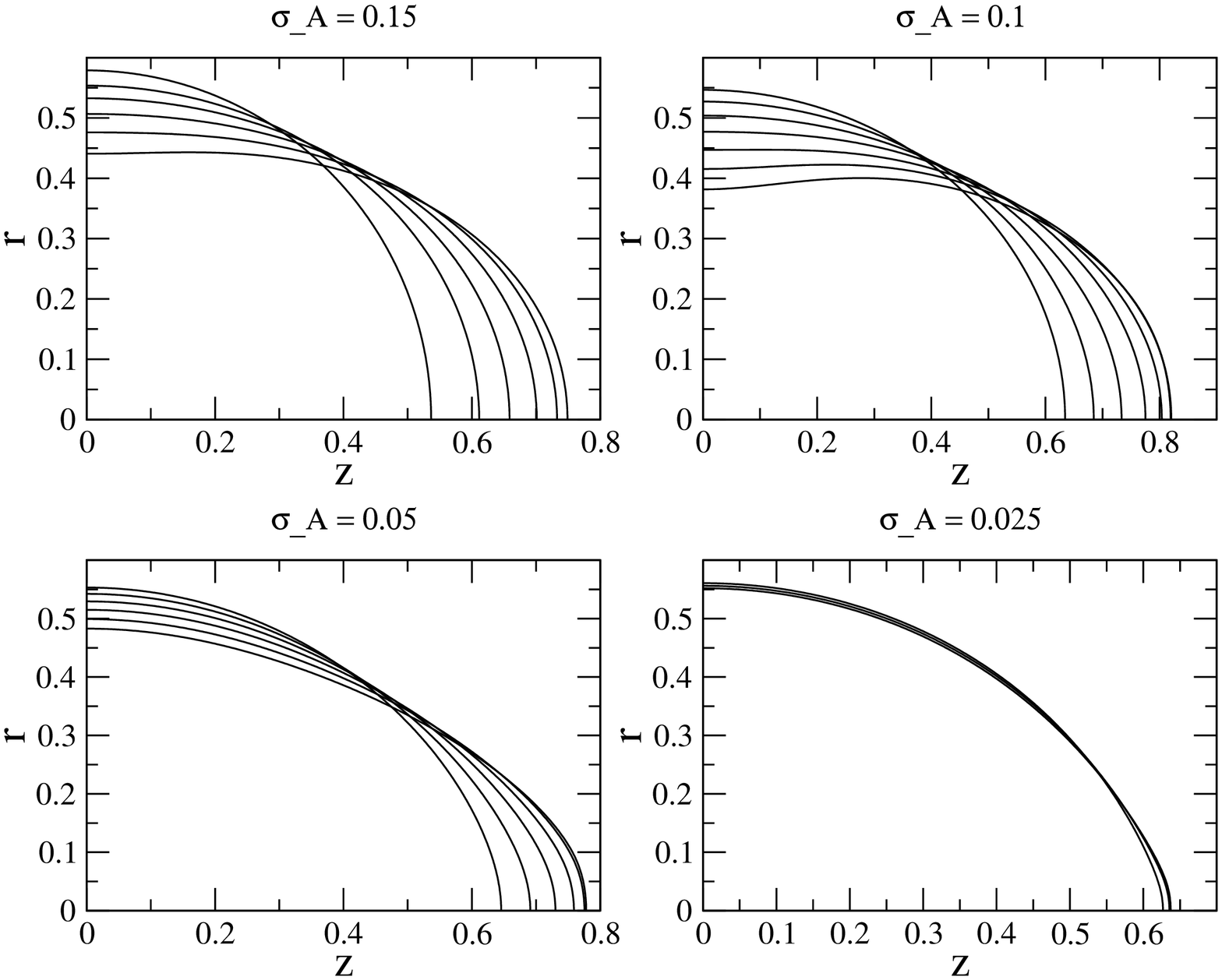,height=17.5cm,angle=270}
\caption{Distorted black hole horizons all with mass 0.303.
The vertical axis is the radial direction, the horizontal axis is
the periodic direction.  The four sequences of initial data used
$\sigma_A$ values of 0.15, 0.1, 0.05, and 0.025;  $\sigma_B$ values
varied from 0.1 to 1.1;  $\sigma_r$ and $\sigma_z$ were held fixed
at 0.1.}
\label{fig:horizons}
\end{center}
\end{figure}

We first considered the behavior of distorted hyperspherical black
holes.  Using the source given by Eq.~\ref{eqn:source}, we fix both
$\sigma_r$ and $\sigma_z$ to be 0.1.  
This choice would
produce a uniform hyperspherical black hole if the second
term of Eq.~\ref{eqn:source} were absent.  We employ this
second term to  generate sequences of initial data by using increasingly
larger $\sigma_B$ values which produce more distorted black 
holes until a black string horizon
results, holding $\sigma_A$ constant.  Four classes of data are
generated, each with a particular value for $\sigma_A$.  The smaller
the $\sigma_A$ choice is, the more distorted the resulting horizon
will appear.  See Figure \ref{fig:horizons}.

A useful quantity to monitor whether a black hole can be distorted
so as to approach a black string is to consider the proper length 
between the two poles, as done in \cite{sorkinpiran}. We thus evaluate
this quantity by considering
\begin{eqnarray}
l = \int^L_{r^{\left(5\right)}} \psi\left(r=0,z\right)dz. 
\label{eqn:distance}
\end{eqnarray}

\begin{figure}
\begin{center}
\epsfig{file=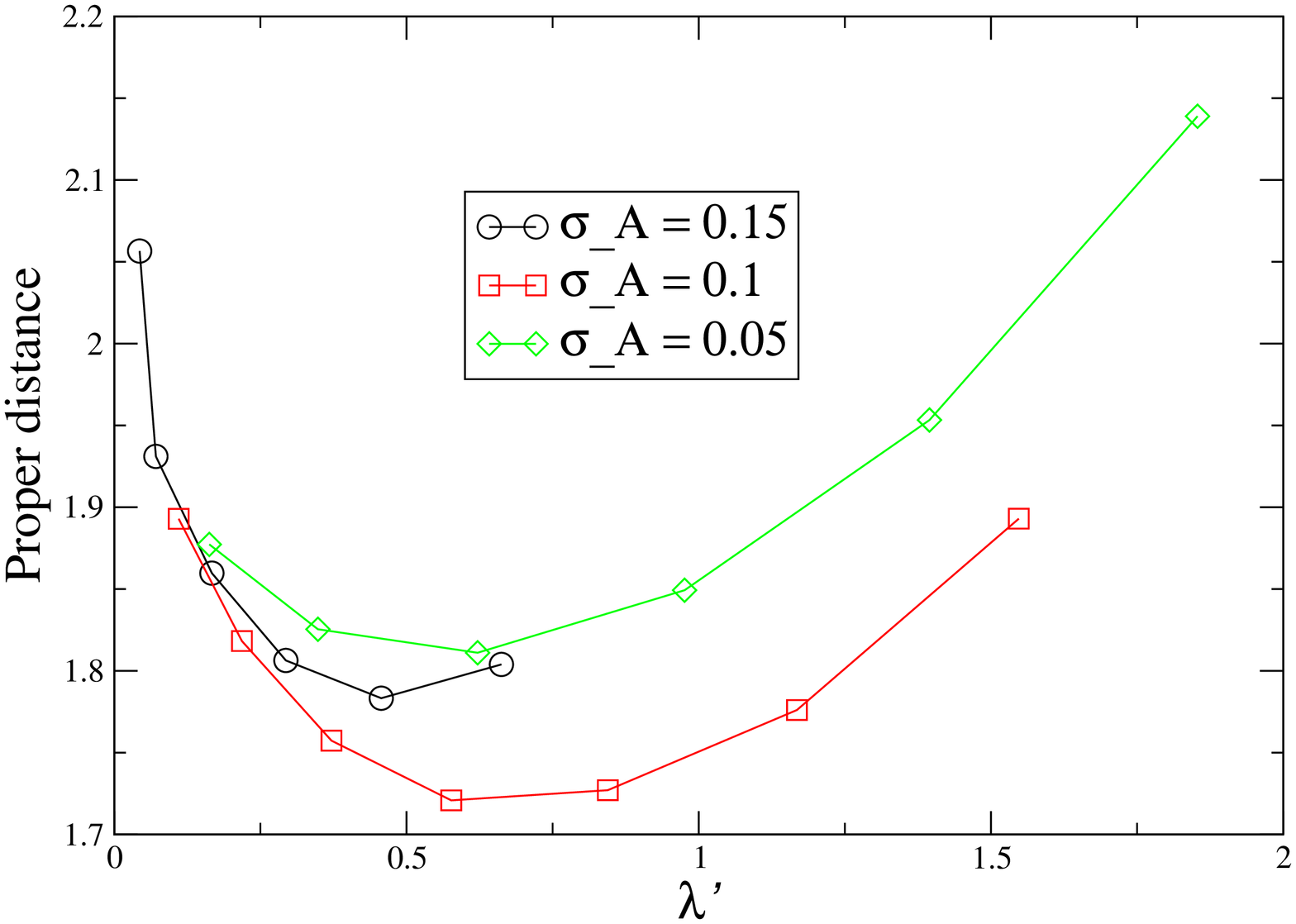,height=12.5cm,angle=270}
\caption{The proper distance versus black hole distortion, $\lambda'$
for the data shown in Figure \ref{fig:horizons}.}
\label{fig:distance}
\end{center}
\end{figure}

For each of the four sequences of initial data, all approach $\sim1.8$
even though the initial data generated using the smallest $\sigma_A$
choices produced more distorted black holes than the largest choices.
The deformation of the hole, $\lambda'$, is given by
\begin{eqnarray}
\lambda' = \frac{R^{\left(5\right)}_{\max}}{R^{\left(5\right)}_{\min}} - 1,
\end{eqnarray}
with $R_{\max},R_{\min}$ the proper distance along the $z$ and $r$ axis.
The results for $l$ for different cases are shown in Figure 
\ref{fig:distance}. Clearly, as indicated by the apparent horizon, 
there does seem to be an
upper limit as to how close the poles can be even when black hole
horizons could be found to be more distorted. This would indicate that
the spacetime in between the poles is becoming considerably curved.


\begin{figure}
\begin{center}
\epsfig{file=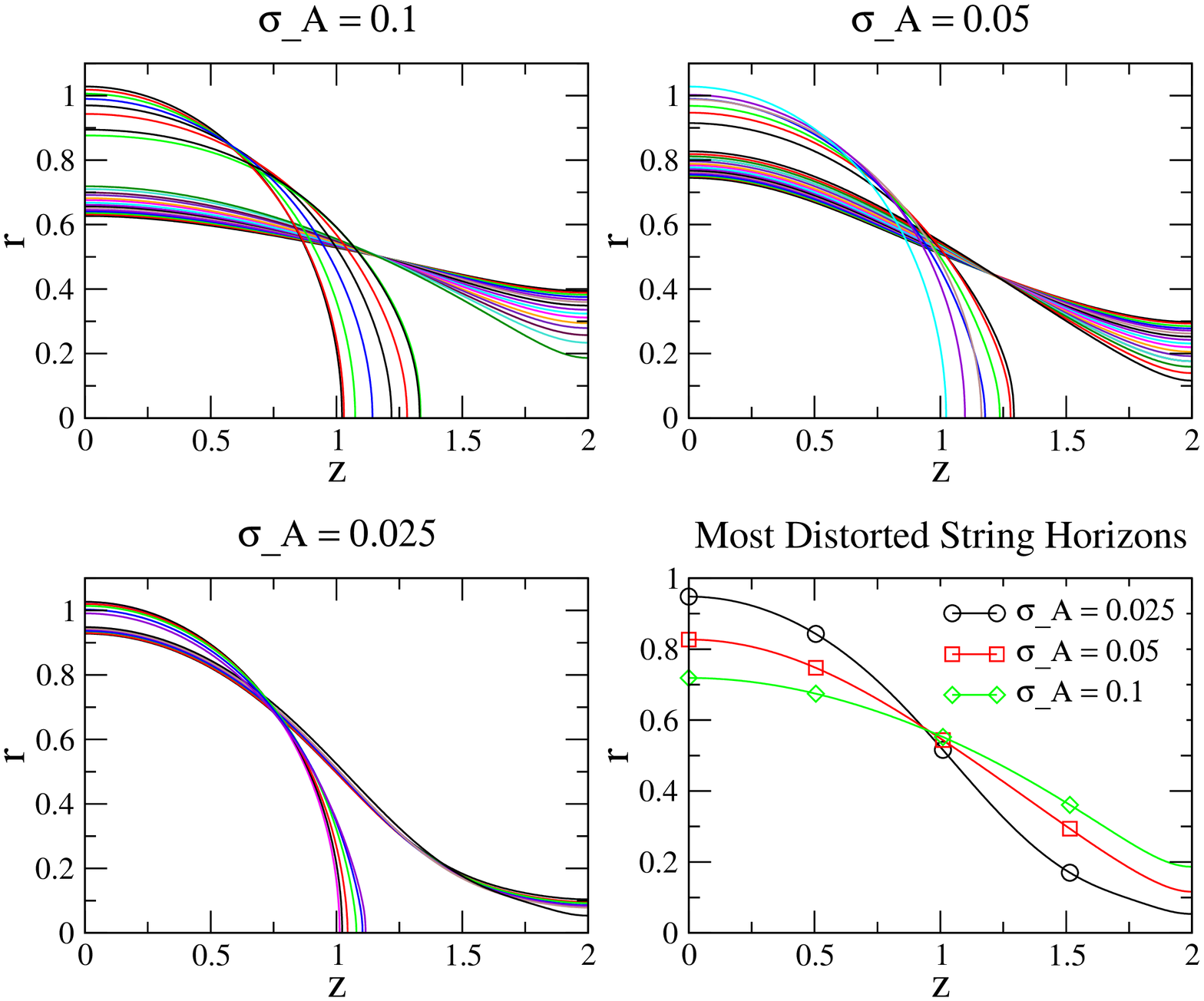,height=17.5cm,angle=270}
\caption{Distorted black hole and black string horizons.
The vertical axis is the radial direction, the horizontal axis is
the periodic direction.  The three sequences of initial data used
$\sigma_A$ values of 0.1, 0.05, and 0.025;  $\sigma_B$ values
varied from 2.0 to 0.1;  $\sigma_r$ and $\sigma_z$ were held fixed
at 0.1.  Smaller choices of $\sigma_A$ produced increasingly distorted
black objects.  The $\sigma_B$ value determines whether a string or hole
is produced.  All data sets had the same ADM mass: $1.52 \pm 0.015$.}
\label{fig:strings}
\end{center}
\end{figure}

\begin{figure}
\begin{center}
\epsfig{file=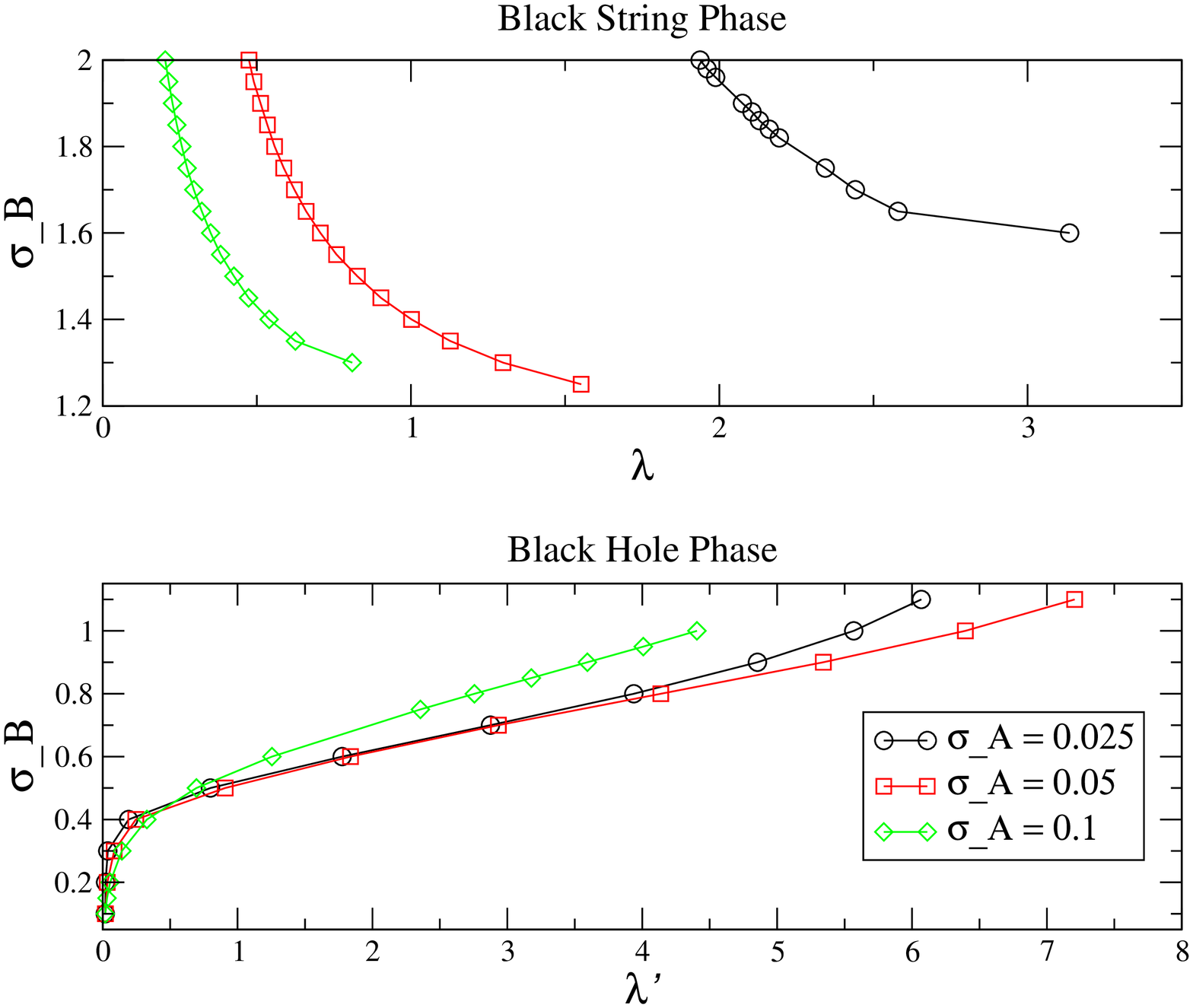,height=17.5cm,angle=270}
\caption{Top: $\sigma_B$ vs. black string phase, $\lambda$;
Bottom: $\sigma_B$ vs. black hole phase, $\lambda'$.
Three sets of $\sigma_A$ values are explored: 0.1, 0.05, and 0.025.  All runs
had the same ADM mass.  As $\sigma_A$ is decreased, increasingly 
distorted black string horizons are found.  See Figure \ref{fig:strings}.
For $\sigma_A = 0.025$, the horizon finder could not distinguish between
a hole or string horizon for $\sigma_B$ values from 1.5 to 1.2}
\label{fig:phase}
\end{center}
\end{figure}

\begin{figure}
\begin{center}
\epsfig{file=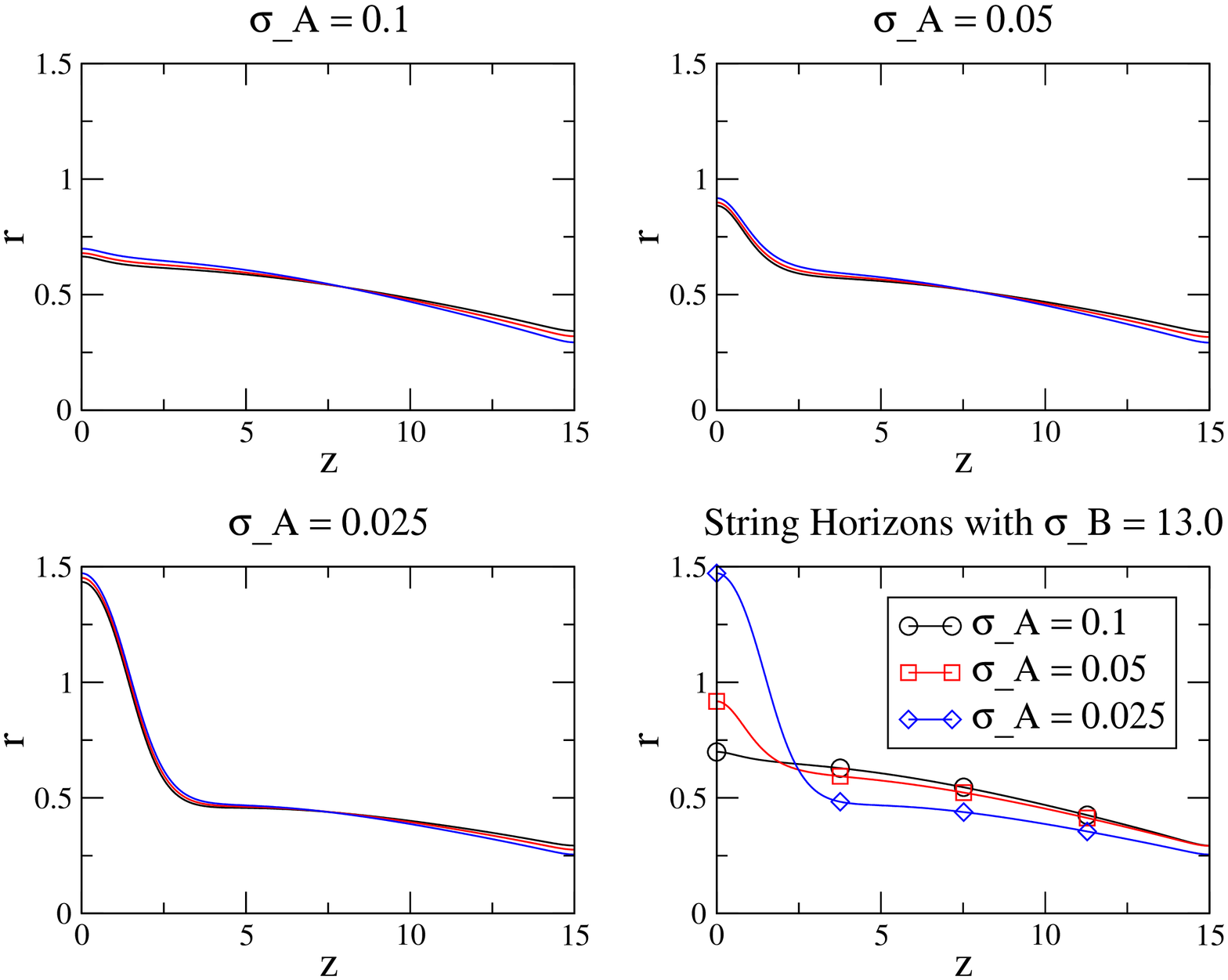,height=17.5cm,angle=270}
\caption{Distorted black string horizons with $L=15$ 
  so as to be above the Gregory--Laflamme critical length.
The vertical axis is the radial direction, the horizontal axis is
the periodic direction.  The three sequences of initial data used
$\sigma_A$ values of 0.1, 0.05, and 0.025;  $\sigma_B$ values
varied from 15.0 to 13.0;  $\sigma_r$ and $\sigma_z$ were held fixed
at 0.1.  All data sets had the same ADM mass: $1.52 \pm 0.015$.
The resolution in these cases was: dr: 0.00225, dz: 0.0145.}
\label{fig:xstring}
\end{center}
\end{figure}

We now turn our attention to a family of gradually more distorted
black strings.
The distortion of a black string, $\lambda$, is given by
\begin{eqnarray}
\lambda = \frac{1}{2} \left( \frac{r^{\left(4\right)}_{\max}}{r^{\left(4\right)}_{\min}} - 1 \right),
\end{eqnarray}
where $r_{\max},r_{\min}$ are the maximum/minimum areal radius
of the apparent horizon.
We fix $\sigma_r=\sigma_z=0.1$ and vary $\sigma_A$ and
$\sigma_B$; this gives rise to black string solutions with large
degree of distortions (see Figure \ref{fig:strings}).  Indeed, we have
found distortions up to $\lambda \simeq 5$ without anything to prevent
obtaining even larger distortions (apart from discretization
issues).  To make this more evident Figure \ref{fig:phase} illustrates
what the distortion $\lambda$ vs $\sigma_B$ is for three different
values of $\sigma_A$ (Recall that $\sigma_A,\sigma_B$ govern the
source's deformation along $r$,$z$ directions).  The same behavior
is evident when the periodic dimension is larger than 
the Gregory--Laflamme critical length (see Fig. \ref{fig:xstring}).

Finally, we examine the Kretschmann invariant $K$ value for different
members of the sequence, in particular we consider the quantity $I=K
R_{ah}^4/12$ which, as mentioned in \cite{Simulation}, would give a
value of $6$ for a uniform hyperspherical black hole while $1$ for a
uniform stationary black string. Interestingly, near the origin, the value of
this invariant is close to $6$ while relatively close to $1$ at $z
\simeq 1$.  Furthermore, the region in between behaves in a way quite
similar to that found at the late stages of the evolution in
\cite{Simulation}.  This could indicate that the slow down observed
there might be a result of the dynamical solution approaching a time
symmetric phase. If this were the case it would give support to the
conjecture that a ``breathing'' scenario might take place
~\cite{horowitzpersonal}.  In this picture the deformed black string
would reach a stage where the trend would reverse itself; that is the
waist of the string would grow while the bulge would shrink. Whether
this takes place would have to be elucidated by numerical evolutions.

\begin{figure}
\begin{center}
\epsfig{file=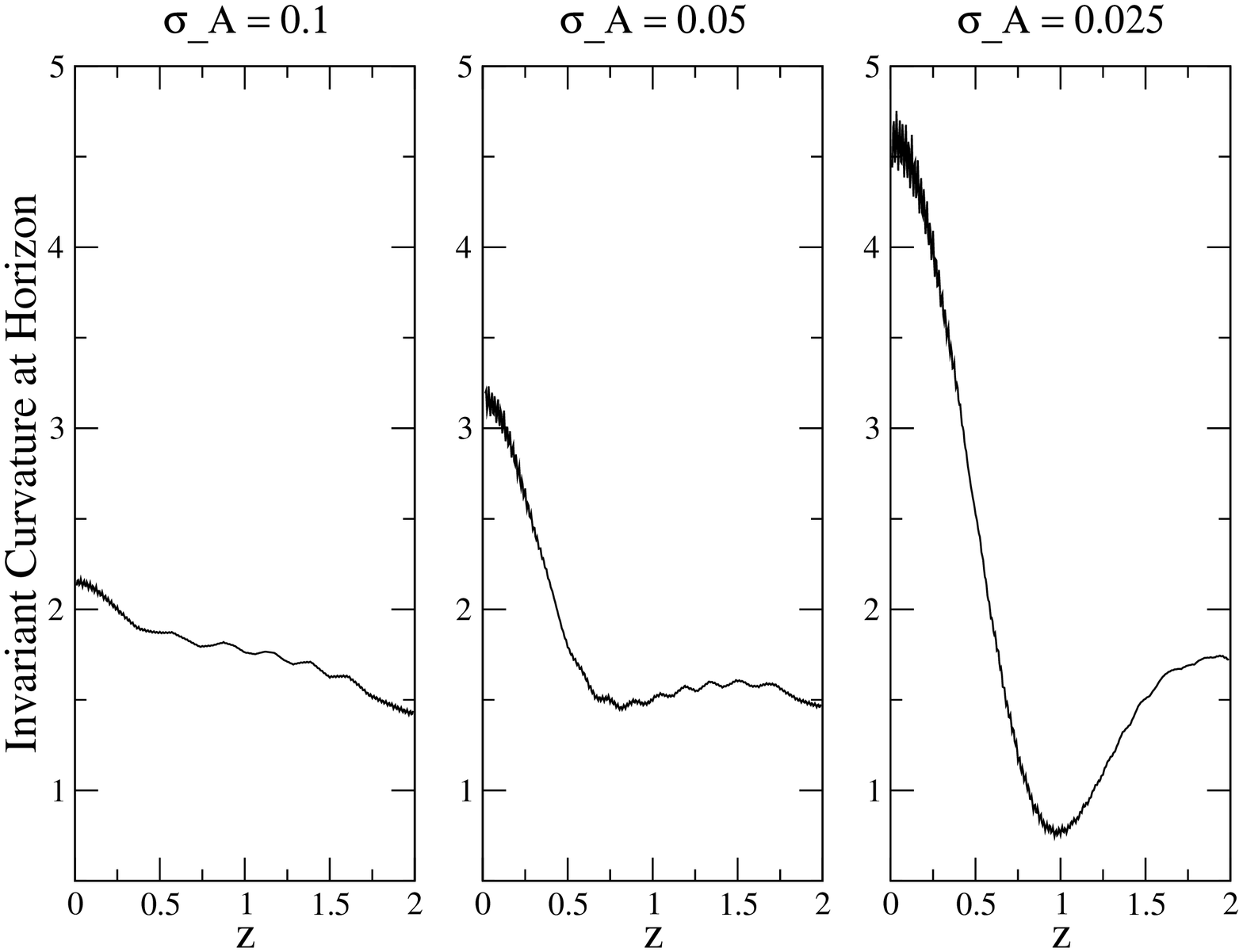,height=17.5cm,angle=270}
\caption{The invariant curvature at the apparent horizon surface
for three sets of distorted black string data using $\sigma_A$ values 
of 0.1, 0.05, and 0.025 and a $\sigma_B$ value of 1.7.  
All simulations had the same ADM mass.
  The curvature appears increasingly closer to
  that for a hyperspherical black hole rather than a black string as the
  black string is more distorted.  See Figures \ref{fig:strings}, \ref{fig:phase}.}
\label{fig:curvature}
\end{center}
\end{figure}

\section{Conclusions}

In the present work we have developed a code based on finite element techniques
capable of providing solutions of the constraints describing highly distorted 
black holes and black strings in spacetimes with arbitrary dimensions $D$
endowed with an $SO(D-2)$  symmetry.
We employed such code to investigate a sequence of solutions which contain a
marginally trapped surface where both the ADM mass and the asymptotic
length of the extra (periodic) dimension are kept fixed. 
We have concentrated on the $D=5$ case to make contact with results presented
in the literature. (The higher dimensional case is under investigation).
The solutions are obtained by providing an arbitrary source (subject to the
requirements of the sequence) which can be varied in different regions so as
to produce highly distorted black holes and string. Our solutions cover both
stable and unstable regimes as far as the the ratio $L/M_{ADM}$ is concerned
and the observed behavior is qualitatively similar in both cases.

A study of the sequences obtained reveals several interesting features.
First, arbitrary distortions
can be obtained both in the black string and black hole phases. Second,
even though we can produce arbitrary distorted black holes, the proper
distance between the poles seem to have an upper limit in how close these
can be made to be as judged by the apparent horizon.
Third, we see no evidence of a conical structure developing at or near
the waist in the apparent horizon region. The spacetime is 
smooth in the horizon's vicinity. Note however that the apparent horizon
would miss a conical structure at the event horizon unless 
the spacetime were stationary.
Last,  the curvature invariant of the apparent horizon corresponding
to highly distorted black strings would indicate that the bulge of the
horizon approaches a hyperspherical black hole while the neck approaches a black string.
Interestingly, this is quite similar to the results obtained at the late
stages of the black string simulations presented in \cite{Simulation}.

\section{Acknowledgments}
This research was supported in part by the NSF under Grants No PHY0244335,
PHY0244699, PHY0326311, INT0204937, NASA-NAG5-13430 and funds from the
Horace Hearne Jr. Laboratory for Theoretical Physics,  CCT-LSU, and a Research
Innovation Award from the Research Corporation to Louisiana State University.
L. L. was partially supported by the Alfred P. Sloan Foundation.

We wish to thank the Albert Einstein Institute and the
National University of Cordoba, Argentina for hospitality
at different stages of this work.

The authors thank F. Pretorius, O. Sarbach, E. Sorkin and T. Wiseman for discussions.

\end{document}